\documentclass[preprintnumbers,nofootinbib,superscriptaddress,10.5pt]{revtex4}
\usepackage{amsmath,amssymb,amsfonts,bm} 

\def\a{\alpha} \def\b{\beta} \def\g{\gamma}  \def\d{\delta} \def\D{\Delta}    \def\th{\theta}     \def\m{\mu}         \def\t{\tau}       

  \def\nn{\nonumber}

\renewcommand{\Im}{{\rm Im}\,}

\newcommand{\Diag}[3]{ \begin{pmatrix} #1 & 0 & 0 \\ 0 & #2 & 0 \\ 0 & 0 & #3 \\\end{pmatrix}}

\usepackage{color}

\begin{document}

\title{\large Rephasing invariant CP phases and sum rules in TM$_{1,2}$ mixing }
\preprint{STUPP-26-298}

\author{Masaki J. S. Yang}
\email{mjsyang@mail.saitama-u.ac.jp}
\affiliation{Department of Physics, Saitama University, 
Shimo-okubo, Sakura-ku, Saitama, 338-8570, Japan}



\begin{abstract} 

We show that the CP phases $\phi_{1,2}$ appearing in the TM$_{1,2}$ mixing are directly identified with rephasing invariants  $\phi _{1} = - \arg \left[  { U_{e2} U_{e3}  U_{\mu 1} U_{\tau 1} / U_{e 1} \det U } \right]$, $\phi_{2}  = - \arg \left[  { U_{e1} U_{e3}  U_{\mu 2} U_{\tau 2} / U_{e 2} \det U} \right]$. 
Furthermore, we demonstrate identities $\phi_{i} = \delta - \arg [ U_{\mu i}^{0} U_{\tau i}^{0}  ]$ among $\phi_{1,2}$, the Dirac CP phase $\delta$ and matrix elements in the PDG parametrization $U^{0}_{\alpha i}$.
These relations of CP phases are interpreted as specific elements of general sum rules 
represented by phase matrices. 

\end{abstract} 

\maketitle

\section{Introduction}

With the progress of neutrino oscillation experiments, the structure of the lepton mixing matrix has been determined with increasing precision. 
Following the discovery of a nonzero $\theta_{13}$ by the Daya Bay experiment \cite{DayaBay:2012fng}, TM$_{1,2}$ mixing \cite{Xing:2002sw, Xing:2006ms, Albright:2008rp,Albright:2010ap,He:2011gb} has been studied \cite{Luhn:2013lkn,Li:2013jya,King:2019vhv,Krishnan:2020xeq, Shimizu:2017fgu, Shimizu:2017vwi, Zhao:2021efc, Jiang:2025hvq, Singh:2026sfi} as a mixing scheme that partially preserves the  structure of the tri-bi-maximal (TBM) mixing \cite{Harrison:2002er}.
The TBM mixing is characterized by a residual $Z_{2}\times Z_{2}$ symmetry \cite{Lam:2006wm,Lam:2007qc,Lam:2008rs}, whereas TM$_{1,2}$ mixings preserve only one of the two $Z_2$ symmetries.
In particular, the residual symmetry associated with TM$_2$ is also referred to as the {\it magic symmetry} \cite{Harrison:2002kp, Lam:2006wy, Friedberg:2006it,Bjorken:2005rm, He:2006qd,Grimus:2008tt, Channey:2018cfj,Hyodo:2020ysq, Bao:2021zwu, Minamizawa:2022fch}.

To describe the TM$_{1,2}$ mixing, parametrizations different from the standard convention can be more suitable.
In general, a $3 \times 3$ unitary mixing matrix can be represented by nine Euler-angle-like parametrizations using three rotations and one complex phase \cite{Fritzsch:1997st, Rasin:1997pn}, 
and these CP phases have been studied through various rephasing invariants \cite{Jarlskog:1985ht, Wu:1985ea, Bernabeu:1986fc, Gronau:1986xb, Branco:1987mj, Bjorken:1987tr, Nieves:1987pp, Botella:1994cs,  Kuo:2005pf,  Farzan:2006vj, Jenkins:2007ip, Branco:2008ai, Esmaili:2007av, Ge:2011qn, Petcov:2014laa, Chiu:2015ega}. 
Recently, by using $\arg \det U$ of the mixing matrix $U$ as a global phase information, 
the CP phases corresponding to each parametrization are directly expressed as an additive argument of matrix elements and the global phase \cite{Yang:2025hex,Yang:2025cya,Yang:2025law, Luo:2025wio,Yang:2025ftl,Yang:2025dhm,Yang:2025vrs}.  
Furthermore, relations among rephasing transformations between different parametrizations have  also been investigated \cite{Yang:2025dkm, Yang:2025qlg, Yang:2026ulu, Yang:2026wjg, Yang:2026pdm}.

In this letter, we demonstrate that the CP phases $\phi_{1,2}$ appearing in the TM$_{1,2}$ mixing are not merely parametrization-dependent quantities, but are exactly identified with specific rephasing invariants associated with the nine Euler-like parametrizations. 
In addition, we derive that the small differences between $\phi_{1,2}$ and the Dirac CP phase $\d$ are written as sums of nontrivial arguments of matrix elements in the Particle Data Group (PDG) parameterization $U^{0}_{\a i }$.  
The differences among CP phases can be understood as sum rules among physical rephasing invariants.

\section{CP phases in nine parametrizations of mixing matrix}

Here, we first review the nine Euler-angle-like parametrizations of the mixing matrix proposed by Fritzsch and Xing \cite{Fritzsch:1997st}, as well as the corresponding formulae of their CP phases \cite{Yang:2025vrs}.
We begin by defining matrices $R_{ij}$ that describe two-dimensional rotations  
\begin{align}
R_{12}(\theta)  =
\begin{pmatrix}
c_{\theta} & s_{\theta} & 0 \\
- s_{\theta} & c_{\theta} & 0 \\
0 & 0 & 1 \\
\end{pmatrix} , ~~~
R_{23}(\sigma)  = 
\begin{pmatrix}
1 & 0 & 0 \\
0 & c_{\sigma} & s_{\sigma} \\
0 & - s_{\sigma} & c_{\sigma} \\
\end{pmatrix} , ~~ 
R_{31} (\tau)  =
\begin{pmatrix}
c_{\tau} & 0 & s_{\tau} \\
0  & 1 & 0 \\
- s_{\tau} & 0 & c_{\tau} \\
\end{pmatrix} , 
\label{Rij}
\end{align}
where $s_{x} = \sin x , c_{x} = \cos x. $ 
Furthermore, complex mixing matrices $R_{12}(\theta, \delta)$, $R_{23}(\sigma, \delta)$, and $R_{31}(\tau, \delta)$ are defined by replacing $1 \rightarrow e^{- i\delta}$ in Eq.~(\ref{Rij}).
Using these rotations, the mixing matrix admits nine distinct parametrizations \cite{Fritzsch:1997st}. 

The two important examples are
\begin{align}
\begin{array}{ll}
P2: ~ U^{(11)} = R_{23}(\sigma) R_{12}(\theta, \d^{(11)}) R^{-1}_{23}(\sigma') \, , &
P7: ~ U^{(12)} = R_{23}(\sigma) R_{12}(\theta, \d^{(12)}) R^{-1}_{31}(\tau) \, .
\end{array}
\end{align}
Here, the parametrization $U^{(\alpha i)}$ and its CP phase $\delta^{(\alpha i)}$ 
are distinguished by matrix elements $U_{\a i}$ with trivial phases satisfying $\arg U_{\alpha i} \in \mathbb{R}$ for a certain row $\alpha$ and column $i$. 
Specifically, 
\begin{align}
\arg U_{\a 1}^{(\a i)} = \arg U_{\a 2}^{(\a i)} = \arg U_{\a 3}^{(\a i)} = \arg U_{ei}^{(\a i)} = \arg U_{\m i }^{(\a i )} = \arg U_{\t i}^{(\a i )} =  0 ~ {\rm or} ~ \pi \, . 
\end{align}
The superscripts $U^{(\a i)}$ distinguish the parameterizations, 
whereas the subscripts $U_{\a i}$ specify the matrix elements. 
Throughout this letter, all expressions containing arg are understood modulo $2\pi$. 
These trivial phases can take the value $\pi$ only for zero or two elements. 
Explicit forms of the first two examples are 
\begin{align}
P2: ~ U^{(e1)} 
& = 
\begin{pmatrix}
c^{~}_{\theta} & s^{~}_{\theta} c_{\sigma^{\prime}} & -s^{~}_{\theta} s_{\sigma^{\prime}} \\ 
-s^{~}_{\theta} c_{\sigma} 	& c^{~}_{\theta} c_{\sigma} c_{\sigma^{\prime}} + s_{\sigma} s_{\sigma^{\prime}} e^{- i \d^{(11)}}	
& -c^{~}_{\theta} c_{\sigma} s_{\sigma^{\prime}} + s_{\sigma} c_{\sigma^{\prime}} e^{- i \d^{(11)} } \\
s^{~}_{\theta} s_{\sigma} & -c^{~}_{\theta} s_{\sigma} c_{\sigma^{\prime}} + c_{\sigma} s_{\sigma^{\prime}} e^{-  i \d^{(11)} }	
& c^{~}_{\theta} s_{\sigma} s_{\sigma^{\prime}} + c_{\sigma} c_{\sigma^{\prime}} e^{-  i \d^{(11)} } 
\end{pmatrix} , \label{P2} \\
P7: ~ U^{(e2)} 
& = 
\begin{pmatrix}
c^{~}_{\theta} c_{\tau} 	& s^{~}_{\theta}  	& -c^{~}_{\theta} s_{\tau} \\
-s^{~}_{\theta} c_{\sigma} c_{\tau} + s_{\sigma} s_{\tau} e^{- i \d^{(12)}}	& c^{~}_{\theta} c_{\sigma}
& s^{~}_{\theta} c_{\sigma} s_{\tau} + s_{\sigma} c_{\tau} e^{- i \d^{(12)}} \\
s^{~}_{\theta} s_{\sigma} c_{\tau} + c_{\sigma} s_{\tau} e^{-i \d^{(12)}}	& -c^{~}_{\theta} s_{\sigma}	
& -s^{~}_{\theta} s_{\sigma} s_{\tau} + c_{\sigma} c_{\tau} e^{- i \d^{(12)}} 
\end{pmatrix} .  \label{P7} 
 \end{align}

The rephasing transformations to parametrizations $U^{(\alpha i)}$ and their CP phases $\d^{(\a i)}$ are  determined by using $\arg \det U$. 
Suppose that unphysical phases of $U$ in an arbitrary basis are factorized by a rephasing transformation, yielding one of the parametrizations $U^{(\alpha i)}$;
\begin{align}
U = \Psi_{L} U^{(\a i )}  \Psi_{R} ^{\dagger} \, , ~~
U_{\b j} = e^{i \g_{L \b}} U^{(\a i)}_{\b j} e^{- i \g_{R j}} \, . 
\end{align}
Here, $(\Psi_{L})_{\a\b} = e^{i \gamma_{L \a}} \d_{\a \b}$ and $(\Psi_{R})_{ij} = e^{i \gamma_{R i}} \d_{ij }$ are diagonal phase matrices. 
Focusing on $\arg  \det U = - \d^{(\a i )} +  \sum_{\b} \g_{L \b} - \sum_{j} \g_{R j }$, 
the CP phases $\d^{(\a i)}$ are solved as \cite{Yang:2025vrs}
\begin{align}
\d^{(\a i )} & =    \sum_{\b} \g_{L \b} -  \sum_{j}  \g_{R j } - \arg  \det U 
 = \arg \left[ { U_{\a 1} U_{\a 2} U_{\a 3} U_{ei} U_{\m i} U_{\t i}  \over U_{\a i }^{3} \det U } \right] . 
\end{align}
Although these matrix elements and the determinant transform under rephasing, 
the resulting CP phases are rephasing invariant because all indices cancel out in the numerator and denominator. 
These quantities can be represented in the complex plane like the unitarity triangles \cite{Yang:2025dhm}, 
and therefore correspond to physical observables. 
Due to the rephasing invariance, the formulae remain valid under changes in the values of the Majorana phases. 

Since the element $U_{\alpha i}$ appears twice in the numerator, 
the phases are represented by arguments of five matrix elements in a certain row and column  
\begin{align}
\d^{(e1)} & = \arg \left[ { U_{e2} U_{e3}  U_{\m 1} U_{\t 1}  \over U_{e 1} \det U } \right]  , ~~ 
\d^{(e2)} = \arg \left[ { U_{e1} U_{e3}  U_{\m 2} U_{\t 2}  \over U_{e 2} \det U} \right] , ~~
\d^{(e3)} = \arg \left[ { U_{e1} U_{e2}  U_{\m 3} U_{\t 3}  \over U_{e 3} \det U} \right]  .  
\label{d123}
\end{align}
Such invariants have also been studied under the constraint $\det U = 1$ and without  $\arg U_{\a i}$
\cite{Kuo:2005pf, Chiu:2015ega}.
Well-known examples are the Dirac CP phase $\delta$ in the Chau--Keung (or PDG) parametrization \cite{Chau:1984fp}, the Kobayashi--Maskawa phase $\delta_{\rm KM}$ in the original  parametrization \cite{Kobayashi:1973fv}, and the phase $\delta_{\rm FX}$ in the  Fritzsch--Xing parametrization \cite{Fritzsch:1997fw}, 
\begin{align}
\d^{(e1)} = \pi - \d_{\rm KM} \, , ~~~ \d^{(e3)} = \d \, , ~~~
\d^{(\t 3)} = \d_{\rm FX} \, . 
\label{dKMPDG}
\end{align}
Since the rephasing invariants take the same values in any basis of phases, 
they coincide with the CP phases defined in the specific conventions.

The equivalence with the Jarlskog invariant $J =\Im [ U_{\a i} U_{\b j} U_{\a j}^{*} U_{\b i}^{*}]$ can be readily shown \cite{Yang:2025hex} 
\begin{align}
\sin  \arg \left[ { U_{e1} U_{e2}  U_{\m 3} U_{\t 3}  \over U_{e 3} \det U} \right] =  
\frac{  1 - |U_{e 3}^{2}| }{ |U_{e 1} U_{e 2} U_{\m 3} U_{\t 3} U_{e 3}| } J 
= \sin \d\, . 
\end{align}
Compared with $J$, this expression has several advantages. 
It can be decomposed as $\arg[ab] = \arg a + \arg b$ which is simpler than the addition theorem of $\sin (a + b)$, 
this phase contains the signature of $\cos \d$, and finally 
its constraint is  independent of those of the mixing angles, 
\begin{align}
 \tan^{2} \th_{12} = {U_{e2}^{*} U_{e2} \over U_{e1}^{*} U_{e1}} \, , ~~
 \tan^{2} \th_{23} = {U_{\m 3}^{*} U_{\m 3} \over U_{\t 3}^{*} U_{\t 3}} \, , ~~ 
 \sin^{2}  \th_{13} = U_{e3}^{*} U_{e3} \, , 
\end{align}
which are defined by the invariant moduli. 

\section{CP phases of lepton mixing matrix and TM$_{1,2}$ mixing}

We now apply this framework to the lepton mixing matrix and show that $\delta^{(e1)}$ and $\delta^{(e2)}$ directly correspond to the CP phases associated with TM$_1$ and TM$_2$ mixing \cite{Xing:2002sw, Xing:2006ms, Albright:2008rp,Albright:2010ap,He:2011gb}.
As the latest global-fit values, we use the results for the normal hierarchy (NH) with Super-Kamiokande (SK) and the inverted hierarchy (IH) without SK \cite{Esteban:2024eli} 
\begin{align}
\begin{array}{llll}
\th_{12}^{\rm NH} = 33.68^{\circ} \, , & \th_{23}^{\rm NH} = 43.3^{\circ} \, , &  \th_{13}^{\rm NH} = 8.56^{\circ} \, , & \d/^{\circ} = 212^{+26}_{-41}  \, ,  \\[2pt]
 \th_{12}^{\rm IH} = 33.68^{\circ} \, , &  \th_{23}^{\rm IH} = 48.6^{\circ} \, , &   \th_{13}^{\rm IH} = 8.58^{\circ}  \, , & \d/^{\circ}= 285^{+25}_{-28}  \, . 
\end{array}
\end{align}
The reason for this choice is that $\theta_{23}$ is very close to $\theta_{23}^{\rm IH}$ in the remaining two cases, 
NH without SK and the IH with SK.
The mixing angles are fixed to their best fits, while the uncertainties of the CP phase are included within the $1\sigma$ range.
For later convenience, we define the phase matrix 
\begin{align}
\D \equiv 
\begin{pmatrix}
\d^{(e1)} & \d^{(e2)} & \d^{(e3)}  \\
\d^{(\m 1)} & \d^{(\m 2)} & \d^{(\m 3)}  \\
\d^{(\t 1)} & \d^{(\t 2)} & \d^{(\t 3)} 
\end{pmatrix} . 
\end{align}
The phases obtained from the global fits are 
\begin{align}
\D^{\rm NH} / ^{\circ} & = 
\begin{pmatrix}
 -33.76 & -32.88 & -148.00 \\
 -163.67 & -13.86 & -37.10 \\
 -17.21 & -167.90 & -29.53 \\
\end{pmatrix}  
+ 
\begin{pmatrix}
^{-25.47}_{+43.42} & ^{-26.20}_{+42.16} & ^{+26}_{-41} \\[2pt]
^{+9.47}_{-21.20} & ^{-7.22}_{+18.06} & ^{-27.94}_{+47.72} \\[2pt]
^{-9.68}_{+22.36} & ^{+7.74}_{-15.64} & ^{-23.74}_{+37.87} \\
\end{pmatrix} , 
\\
\D^{\rm IH} / ^{\circ} & = 
\begin{pmatrix}
-106.60 & -103.33 & -75.00 \\
-149.28 & -24.10 & -111.56 \\
-29.05 & -157.51 & -98.37 
\end{pmatrix} 
+ 
\begin{pmatrix}
^{-23.07}_{+25.33} & ^{-24.99}_{+27.47} & ^{+25}_{-28} \\[2pt]
^{-6.08}_{+0.53} & ^{+5.64}_{-1.70} & ^{-22.61}_{+25.97} \\[2pt]
^{+6.10}_{-1.06} & ^{-3.71}_{-0.97} & ^{-25.44}_{26.83}
\end{pmatrix} .
\label{Dfit}
\end{align}
Here, the branch of arg is chosen to be $(-\pi , \pi]$.  

We proceed to analyze the following TM$_{1,2}$ mixings
\begin{align}
U_{\rm TM1} & = U_{\rm TBM} U_{23} 
=  
\begin{pmatrix}
 \sqrt{\frac{2}{3}} & \frac{c_{1}}{\sqrt{3}} & \frac{1 }{\sqrt{3}} s_{1} e^{-i \phi _1} \\
 -\frac{1}{\sqrt{6}} & \frac{c_{1}}{\sqrt{3}}+\frac{s_{1} }{\sqrt{2}} e^{i \phi _1} & -\frac{c_{1}}{\sqrt{2}}+\frac{s_{1} }{\sqrt{3}} e^{-i \phi _1}\\
 -\frac{1}{\sqrt{6}} & \frac{c_{1}}{\sqrt{3}}-\frac{s_{1} }{\sqrt{2}} e^{i \phi _1} & \frac{c_{1}}{\sqrt{2}}+\frac{s_{1} }{\sqrt{3}} e^{-i \phi _1}\\
\end{pmatrix}  ,   \label{TM1}
\\
U_{\rm TM2} & = U_{\rm TBM} U_{13} 
=
\begin{pmatrix}
 \sqrt{\frac{2}{3}} c_{2} & \frac{1}{\sqrt{3}} & \sqrt{\frac{2}{3}} s_{2} e^{-i \phi _2} \\
 -\frac{c_{2}}{\sqrt{6}}+\frac{s_{2} }{\sqrt{2}} e^{i \phi _2} & \frac{1}{\sqrt{3}} & -\frac{c_{2}}{\sqrt{2}}-\frac{s_{2} }{\sqrt{6}} e^{i \phi _2} \\
 -\frac{c_{2}}{\sqrt{6}}-\frac{s_{2} }{\sqrt{2}} e^{i \phi _2} & \frac{1}{\sqrt{3}} & \frac{c_{2}}{\sqrt{2}}-\frac{s_{2} }{\sqrt{6}} e^{i \phi _2}\\
\end{pmatrix} ,  \label{TM2}
\end{align}
where $U_{13}$, $U_{23}$, and  the TBM mixing  with unit determinant $U_{\rm TBM}$  are 
\begin{align}
U_{\rm TBM} 
= 
\begin{pmatrix}
 \sqrt{\frac{2}{3}} & \frac{1}{\sqrt{3}} & 0 \\
 -\frac{1}{\sqrt{6}} &\frac{1}{\sqrt{3}} & - \frac{1}{\sqrt{2}} \\
 -\frac{1}{\sqrt{6}} & \frac{1}{\sqrt{3}} & \frac{1}{\sqrt{2}}  \\
\end{pmatrix} , 
~~~
U_{23} = 
\begin{pmatrix}
1 & 0 & 0 \\
0 & c_{1} & s_{1} e^{- i\phi_{1}} \\
0 & - s_{1} e^{i \phi_{1}} & c_{1}
\end{pmatrix} ,
~~~ 
U_{13} =
\begin{pmatrix}
c_{2} & 0 & s_{2} e^{- i \phi_{2}} \\
0 & 1 & 0 \\
- s_{2} e^{i \phi_{2}} & 0 & c_{2}
\end{pmatrix} .
\label{U123}
\end{align}
From this, the 1-3 mixing $s_{13}$ in the PDG parametrization is 
evaluated from a rephasing invariant 
\begin{align}
{\rm TM}_{1} : s_{13}^{2} = {1\over 3} s_{1}^{2} \, , ~~~ 
{\rm TM}_{2} : s_{13}^{2} =  {2 \over 3} s_{2}^{2}  \, . 
\label{s13}
\end{align}
The Jarlskog invariant is evaluated as 
\begin{align}
J_{\rm TM1} = { c_{1} s_{1} \sin \phi_{1}  \over 3 \sqrt{6}} \, , ~~~ 
J_{\rm TM2} = {c_{2} s_{2} \sin \phi_{2} \over 3 \sqrt 3} \, . 
\end{align}

By contrast, 
these CP phases are evaluated more directly from the rephasing invariant formulae.
From Eq.~(\ref{d123}), the Dirac CP phases (in the PDG parametrization) $\d_{\rm TM1,2} 
\equiv \arg [U_{e1} U_{e2} U_{\m3} U_{\t 3} / U_{e3} \det U_{\rm TM1,2}]$ in TM$_{1,2}$ mixing are found to be
\begin{align}
\d_{\rm TM1} = \phi _1 +  \arg \left[   \frac{s^{2} }{3} e^{- 2 i \phi _1} -\frac{c^{2}}{2}   \right ]  \, , 
~~
\d_{\rm TM2} = \phi _2 + \arg \left[ \frac{s^{2} }{6} e^{- 2 i \phi _2} -  \frac{c^{2}}{2} \right ]  \, .  
\label{phiTM12}
\end{align}
Note that overall real factors do not affect the phases.
From Eq.~(\ref{s13}) and $ s_{13}^2 \simeq 0.02$, 
the parameter $s^2$ is sufficiently small, and these phases $\d_{\rm TM1,2}$ and $\phi_{1,2}$ are  approximately equal up to $\pi$
\begin{align}
\d_{\rm TM1} = \phi _1 + \pi  + O(s_{13}^{2}) \, , ~~~ \d_{\rm TM2} = \phi _2 + \pi + O(s_{13}^{2})\, ,  ~~ ( {\rm mod} ~ 2 \pi) \, . 
\label{approx}
\end{align}

Furthermore, from the parametrizations~(\ref{TM1}) and (\ref{TM2}), 
the phases $\phi_{1,2}$ are directly expressed in terms of rephasing invariants as
\begin{align}
 \d^{(e1)} =  \arg \left[  { U_{e2} U_{e3}  U_{\m 1} U_{\t 1}  \over U_{e 1} \det U } \right] = - \phi _{1} \, , ~~~
\d^{(e2)} =  \arg \left[  { U_{e1} U_{e3}  U_{\m 2} U_{\t 2}  \over U_{e 2} \det U} \right] = - \phi_{2} \, . 
\end{align}
These results are also confirmed by rephasing transformations.  
Decomposing the TBM mixing into two rotations, 
the TM$_{1,2}$ mixings become
\begin{align}
U_{\rm TM1} 
 & = 
\begin{pmatrix}
1 & 0 & 0 \\
0 &\frac{1}{\sqrt{2}} & - \frac{1}{\sqrt{2}} \\
0 & \frac{1}{\sqrt{2}} &  \frac{1}{\sqrt{2}} \\
\end{pmatrix} 
\begin{pmatrix}
 \sqrt{\frac{2}{3}} & \frac{1}{\sqrt{3}} & 0 \\
 -\frac{1}{\sqrt{3}}  &\frac{2}{\sqrt{3}} & 0 \\
0 & 0 & 1\\
\end{pmatrix} 
\begin{pmatrix}
1 & 0 & 0 \\
0 & c_{\th} & s_{\th} e^{- i\phi_{1}} \\
0 & - s_{\th} e^{i \phi_{1}} & c_{\th}
\end{pmatrix} , \\
U_{\rm TM2} & =
\begin{pmatrix}
1 & 0 & 0 \\
0 &\frac{1}{\sqrt{2}} & - \frac{1}{\sqrt{2}}  \\
0 & \frac{1}{\sqrt{2}} &  \frac{1}{\sqrt{2}} \\
\end{pmatrix} 
\begin{pmatrix}
 \sqrt{\frac{2}{3}} & \frac{1}{\sqrt{3}} & 0 \\
 -\frac{1}{\sqrt{3}}  &\frac{2}{\sqrt{3}} & 0 \\
0 & 0 & 1\\
\end{pmatrix} 
\begin{pmatrix}
c_{\th} & 0 & s_{\th} e^{- i \phi_{2}} \\
0 & 1 & 0 \\
- s_{\th} e^{i \phi_{2}} & 0 & c_{\th}
\end{pmatrix} .
\end{align}
These parametrizations are reduced to Eqs.~(\ref{P2}) and (\ref{P7}) with $\phi_i = -\delta^{(1i)}$ by simple rephasing transformations. 
Therefore, the CP phases derived from the rephasing invariant formulae
 are equivalent to those obtained from the TM$_{1,2}$ constraints. 
From the global fit~(\ref{Dfit}),  the TM$_{1,2}$ phases in the IH case can be interpreted as being nearly maximal $\phi_{1,2} \simeq \pi/2$. 

The direct expressions of the CP phases make analyses of the neutrinoless double beta decay more transparent.
Introducing Majorana-like phases $\alpha_{2,3}$ and the neutrino masses $m_i$, the effective mass $m_{ee}$ is given by
\begin{align}
{\rm TM}_{1} : m_{ee} &=  {\frac{2}{3}} m_{1} + \frac{c^{2}}{3} e^{i \a_{2}} m_{2} +
 \frac{1 }{3} s^{2} e^{- 2 i \phi _1 + i \a _{3}}  m_{3} \, , 
\\
{\rm TM}_{2} : 
m_{ee} &= \frac{2}{3} c^{2} m_{1} + \frac{1}{3} e^{i \a_{2}} m_{2} +
\frac{2}{3} s^{2} e^{- 2 i \phi _2 + i \a_{3}} m_{3} \,  ,
\end{align}
and correlations between $\phi_{1,2}$ and $m_{ee}$ become much clearer than the Dirac phase $\delta$.

\subsection{CP phase differences and sum rules with third-order invariants}

Since Eq.~(\ref{approx}) suggests that the phase differences are small, we now derive their exact form expressed in terms of rephasing invariants 
\begin{align}
\d_{\rm TM1} - \phi _1 - \pi & =  \arg \left[  { U_{e1} U_{e2}  U_{\m 3} U_{\t 3}  \over U_{e 3} \det U } \right]+  \arg \left[ -   { U_{e2} U_{e3}  U_{\m 1} U_{\t 1}  \over U_{e 1} \det U } \right] \nn \\
& =  \arg \left[  {  U_{e2}  U_{\m 3} U_{\t 1}  \over \det U } \right]+  \arg \left[ -   { U_{e2}   U_{\m 1} U_{\t 3}  \over  \det U } \right]  = \chi_{2} + \psi_{2} \, , \\
\d_{\rm TM2} - \phi _2 - \pi & =  \arg \left[  { U_{e1} U_{e2}  U_{\m 3} U_{\t 3}  \over U_{e 3} \det U } \right]+  \arg \left[ - { U_{e1} U_{e3}  U_{\m 2} U_{\t 2}  \over U_{e 2} \det U} \right] \nn \\
& = \arg \left[ - { U_{e1}  U_{\m 3} U_{\t 2} \over \det U } \right]+  \arg \left[  { U_{e1}  U_{\m 2} U_{\t 3}  \over \det U} \right] = \chi_{1} + \psi_{1} \, . 
\end{align}
Here, the arguments of the third-order rephasing invariants are defined as \cite{Yang:2025law}
\begin{align}
\chi_{1} = \arg \left[ \frac{ U_{e1} U_{\m 2} U_{\t 3} }{  \det U } \right]  \, , ~
\chi_{2} &= \arg \left[ \frac{  U_{e2} U_{\m 3} U_{\t 1} }{  \det U } \right] \, , ~ 
\chi_{3} = \arg \left[ \frac{  U_{e3} U_{\m 1} U_{\t 2} }{ \det U } \right] \, ,  \nn \\
\psi_{1} = \arg \left[ - \frac{U_{e1} U_{\m 3} U_{\t 2} }{  \det U } \right]  \, ,  ~
\psi_{2} &= \arg \left[ - \frac{  U_{e2} U_{\m 1} U_{\t 3} }{ \det U } \right] \, , ~
\psi_{3} =  \arg \left[ - \frac{  U_{e3} U_{\m 2} U_{\t 1} }{ \det U } \right] \, .  \nn
\end{align}
For later convenience, we assign positive (negative) signs to even (odd) permutations.
These quantities correspond to the four nontrivial arguments of matrix elements 
in the PDG parametrization $U^{0}$,
\begin{align}
\begin{array}{lll}
\chi_{1} = \arg [ U_{\m 2}^{0} ] \, , &
\chi_{2} = \arg [U_{\t 1}^{0} ]  \, , &
\chi_{3} = \arg [  U_{e 3}^{0} U_{\m 1}^{0} U_{\t 2}^{0} ] \, ,  \nn \\[2pt]
\psi_{1} = \arg [ - U_{\t 2}^{0} ]  \, ,  &
\psi_{2} = \arg [ - U_{\m 1}^{0} ] \, , &
\psi_{3} =  \arg [ -  U_{e3}^{0} U_{\m 2}^{0} U_{\t 1}^{0} ] \, ,
\end{array}
\end{align}
because the elements $U^{0}_{e1}$, $U^{0}_{e2}$, $U^{0}_{\mu 3}$, $U^{0}_{\tau 3}$, as well as $\det U^{0}$ have trivial arguments.
Using this notation, the differences ultimately reduce to the following compact identities
\begin{align}
  \phi _1 &  = \d_{\rm TM1} -  \arg [ U_{\m 1}^{0} U_{\t 1}^{0}  ]  \, , ~~~
  \phi _2  = \d_{\rm TM2} - \arg [ U_{\m 2}^{0} U_{\t 2}^{0}  ]  \, .
\end{align}

These relations are also verified by the explicit rephasing transformation from a general mixing matrix $U$ to the PDG parametrization  \cite{Yang:2025dkm} 
\begin{align}
U^{0} = \Diag{  e^{ - i \arg U_{e1} }  } {e^{ i  \arg [ {U_{e2}  U_{\tau 3} \over \det U  } ]}}{ e^{ i \arg [ {  U_{e2} U_{\mu 3} \over \det U } ] } }  U \Diag{ 1} { e^{ - i \arg [{U_{e2} \over U_{e1} } ] }  }
{e^{ - i \arg [ { U_{e2} U_{\mu 3} U_{\tau 3} \over \det U } ] }} . 
\end{align}
Since the parametrizations~(\ref{TM1}) and (\ref{TM2}) have trivial phases $\arg U_{e1} = \arg U_{e2} = \arg \det U = 0$, the rephasing transformation is simplified as 
\begin{align}
U^{0}_{\rm TM1,2} = {\rm diag} (  1 \, ,\,  e^{ i  \arg  U_{\tau 3} } \, , \,  e^{ i \arg  U_{\mu 3}  }  ) \, U_{\rm TM1,2} \, {\rm diag} ( 1 \, , \, 1 \, ,  \, e^{ - i \arg [  U_{\mu 3} U_{\tau 3}  ] }  ) \, . 
\end{align}
From this,  we obtain $\arg [U_{\mu 1}^{0} U_{\tau 1}^{0}] = \arg [U_{\mu 3} U_{\tau 3}]$ in TM$_{1}$ and $\arg [U_{\mu 2}^{0} U_{\tau 2}^{0}] = \arg [U_{\mu 3} U_{\tau 3}]$  in TM$_{2}$.
By comparing the 1-3 elements, one can verify the relation $\delta = \phi_i + \arg [U_{\mu i}^{0} U_{\tau i}^{0}]$.
These sum rules can also be interpreted as transformation laws between the CP phases. The inverse transformations are given by Eq.~(\ref{phiTM12}). 

The best-fit values for the NH and IH cases are 
\begin{align}
\begin{array}{lll}
\chi_{1}^{\rm NH}  =  2.63^{\circ} \, , 
& \chi_{2}^{\rm NH}  =  5.97^{\circ} \, , 
& \chi_{3}^{\rm NH}  = 136.76^{\circ} \, ,  \nn \\[2pt]
\psi_{1}^{\rm NH}  = -3.51^{\circ} \, ,  
& \psi_{2}^{\rm NH}  = -7.73^{\circ} \, , 
& \psi_{3}^{\rm NH}  =  -23.40^{\circ} \, ,  \nn
\end{array}
\end{align}
and 
\begin{align}
\begin{array}{lll}
\chi_{1}^{\rm IH}  =  6.40^{\circ} \, , &
\chi_{2}^{\rm IH}  =  11.36^{\circ} \, ,  &
\chi_{3}^{\rm IH}  = 57.31^{\circ} \, ,  \nn \\[2pt]
\psi_{1}^{\rm IH}  = -4.73^{\circ} \, ,  &
\psi_{2}^{\rm IH}  = -12.96^{\circ} \, ,  &
\psi_{3}^{\rm IH}  =  -87.24^{\circ} \, .  \nn
\end{array}
\end{align}
In this way, the sign is preserved for odd and even permutations.
From these values, the validity of the relations can also be confirmed numerically, 
\begin{align}
{\rm NH} : \d_{\rm TM1} - \phi _1 - \pi  & = -1.76^{\circ} = \chi_{2} + \psi_{2} \, ,  ~~~
\d_{\rm TM2} - \phi _2 - \pi  = -0.88^{\circ} = \chi_{1} + \psi_{1} \, , \\
{\rm IH} : \d_{\rm TM1} - \phi _1 - \pi  & =-1.60^{\circ} = \chi_{2} + \psi_{2} \, ,  ~~~
\d_{\rm TM2} - \phi _2 - \pi  = +1.67^{\circ} = \chi_{1} + \psi_{1}  \, . 
\end{align}

Motivated by the TM$_{1,2}$ relations, we explore general matrix forms that simultaneously encode all nine phase-difference sum rules. 
Defining the following matrices, 
\begin{align}
{\rm X } =
\begin{pmatrix}
\chi_{1} & \chi_{2} & \chi_{3}    \\
\chi_{3} & \chi_{1} & \chi_{2}  \\
\chi_{2} & \chi_{3} & \chi_{1} \\
\end{pmatrix} ,  ~~
\Psi = 
\begin{pmatrix}
\psi_{1} & \psi_{2} & \psi_{3} \\
\psi_{2} & \psi_{3} & \psi_{1} \\
 \psi_{3} & \psi_{1} & \psi_{2} \\
\end{pmatrix} ,  ~~
T = 
\begin{pmatrix}
 0 & 0 & 1 \\
 1 & 0 & 0 \\
 0 & 1 & 0 \\
\end{pmatrix} , ~~
\Pi = 
\begin{pmatrix}
 1 & 1 & 1 \\
 1 & 1 & 1 \\
 1 & 1 & 1 \\
\end{pmatrix} \pi \, ,
\end{align}
we obtain
\begin{align}
\D + \D T^{2} - \Pi &= \arg  \left[  {-1 \over \det U^{2}}
\begin{pmatrix}
 U_{e2}^2 U_{\m1} U_{\m3} U_{\t1} U_{\t3} & U_{e3}^2 U_{\m1} U_{\m2} U_{\t1} U_{\t2} & U_{e1}^2 U_{\m2} U_{\m3} U_{\t2} U_{\t3} \\
 U_{e1} U_{e3} U_{\m2}^2 U_{\t1} U_{\t3} & U_{e1} U_{e2} U_{\m3}^2 U_{\t1} U_{\t2} & U_{e2} U_{e3} U_{\m1}^2 U_{\t2} U_{\t3} \\
 U_{e1} U_{e3} U_{\m1} U_{\m3} U_{\t2}^2 & U_{e1} U_{e2} U_{\m1} U_{\m2} U_{\t3}^2 & U_{e2} U_{e3} U_{\m2} U_{\m3} U_{\t1}^2 \\
\end{pmatrix} \right ]  \nn \\
&= \Psi T   +  {\rm X} T  \, , 
\end{align}
where $\arg$ is applied element-wise to the matrix. 
The 1-1 element corresponds to the TM$_1$ relation $\delta^{(11)} + \delta^{(13)} - \pi = \psi_2 + \chi_2$, 
and the 1-3 element corresponds to the TM$_2$ relation $\delta^{(13)} + \delta^{(12)} - \pi = \psi_1 + \chi_1$. 
Together with the independent relation for columns 
\begin{align}
\D + T \D  -  \Pi =   T^{2} \Psi  + T^{2} {\rm X} \, , 
\end{align}
these two sets of identities are understood as alternative representations of general sum rules that were not discussed in Ref.~\cite{Yang:2025cya}. 
Although the rephasing-invariant expressions are independent of the phase convention, their identification with specific parameters such as $\d$ and $\phi_{1,2}$ depends on the chosen parametrization. 

\section{Summary}

In this letter, we show that the CP phases $\phi_{1,2}$ appearing in the TM$_{1,2}$ mixing are directly identified with rephasing invariants  $\phi _{1} = - \arg \left[  { U_{e2} U_{e3}  U_{\m 1} U_{\t 1} / U_{e 1} \det U } \right]$ , $\phi_{2}  = - \arg \left[  { U_{e1} U_{e3}  U_{\m 2} U_{\t 2} / U_{e 2} \det U} \right]$. 
From the latest global fit, the TM$_{1,2}$ phases become nearly maximal in the inverted hierarchy case.
Furthermore, we demonstrate identities $\phi_{i} = \d - \arg [ U_{\m i}^{0} U_{\t i}^{0}]$ among $\phi_{1,2}$, the Dirac CP phase $\delta$ and matrix elements in the PDG parametrization $U^{0}_{\a i }$. 
These relations  of CP phases are interpreted as specific elements of general sum rules represented by phase matrices.
This approach is also expected to be effective for other residual symmetries \cite{Novichkov:2018yse, Ge:2025csr, Dutta:2026dzh} and  the effective mass of the neutrinoless double beta decay.  



\end{document}